\documentclass[manuscript,screen,review]{acmart}
\AtBeginDocument{%
  \providecommand\BibTeX{{%
    \normalfont B\kern-0.5em{\scshape i\kern-0.25em b}\kern-0.8em\TeX}}}
\copyrightyear{2021}
\acmYear{2021}
\setcopyright{acmlicensed}\acmConference[ICGJ 2021]{Sixth Annual International Conference on Game Jams, Hackathons, and Game Creation Events}{August 2, 2021}{Montreal, Canada}
\acmBooktitle{Sixth Annual International Conference on Game Jams, Hackathons, and Game Creation Events (ICGJ 2021), August 2, 2021, Montreal, Canada}
\acmPrice{15.00}
\acmDOI{10.1145/3472688.3472695}
\acmISBN{978-1-4503-8417-9/21/08}
\usepackage{graphicx}
\begin{document}

\title{Piloting a Game Jam in Nigeria to Support Empathy and Compassion}
\author{Karen Schrier}
\email{kschrier@gmail.com}
\orcid{0000-0003-4114-488X}
\affiliation{%
  \institution{Marist College}
  \country{USA}
}
\author{Eugene A Ohu}
\email{eohu@lbs.edu.ng}
\orcid{0000-0003-1135-267X}
\affiliation{%
  \institution{Lagos Business School}
  \country{Nigeria}
}
\author{Ikeola J Bodunde}
\email{ibodunde@lbs.edu.ng}
\orcid{0000-0002-6128-7565}
\affiliation{%
  \institution{Lagos Business School}
  \country{Nigeria}
}
\author{Morenike O Alugo}
\email{malugo@lbs.edu.ng}
\orcid{0000-0003-1256-7859}
\affiliation{%
  \institution{Lagos Business School}
  \country{Nigeria}
}
\author{Cynthia U Emami}
\email{cemami@lbs.edu.ng}
\orcid{0000-0003-0820-0886}
\affiliation{%
  \institution{Lagos Business School}
  \country{Nigeria}
}
\author{Esther O A Babatunde}
\email{ababatunde@lbs.edu.ng}
\orcid{0000-0001-8983-6266}
\affiliation{%
  \institution{Lagos Business School}
  \country{Nigeria}
}
\renewcommand{\shortauthors}{Schrier, et al.}
\begin{abstract}
While games may help to support skills practice and attitudinal change, the game creation process itself may also be effective in enhancing empathy and compassion for other people. Nigeria has over 250 ethnic groups and 500 languages amid its 200 million people. We aim to enhance perspective-taking, empathy, and compassion across different ethnic groups using game jams in public schools in Nigeria. In 2021, we piloted a game jam focused on identity exploration and perspective-taking for students ages 12 to 20. Initial results and next steps are shared.
\end{abstract}
\maketitle
\section{Introduction}
Nigeria, a country with over 200 million people, has a largely Muslim north and a Christian south, more than 250 ethnic minority groups, and more than 500 languages. Tribalism causes some Nigerians to show concern for their own ethnic group while being indifferent to other groups, even when these other groups are suffering. Such tensions were partly responsible for the bitter genocidal civil war Nigeria fought from 1967 to 1970, where six million people died, out of which one million children from eastern Nigeria died of hunger. Violence and tensions have continued, and though poverty and disease are worse in some parts of Nigeria, it is met with little compassion by the rest of the country.

Youth are a significant proportion of Nigeria’s population, with over 60\% of the people under age 18. Our goal is to enhance the practice of empathy and compassion, and related skills, for adolescents ages 12 to 20. One way to encourage the development and practice of these skills is through participation in game design and development, such as through game jam events \cite{schrier2019designing}. Game jams are short, intense periods of game creation, and often occur over the course of a short period of time, such as a week or weekend, which may happen in person, online, or even in multiple cities or countries \cite{schrier2019designing},\cite{fowler2020a},\cite{kultima2015a}. We aim to create and deploy a series of game jam experiences to youth in Nigerian public schools, and to investigate their efficacy in supporting the practice of empathy and compassion skills (such as perspective-taking), and encouraging behavior and attitudinal change, such as compassion toward ethnic groups that are not one’s own.
\subsection{Empathy, compassion and games}

Empathy is the ability to understand and share the feelings of others. It often includes affective, cognitive, and motivational components \cite{schrier2021a}\cite{schrier2021b}. Cognitive empathy describes purposefully taking on someone else’s point of view \cite{belman2010a}, and affective empathy typically connects empathy to emotions and feelings \cite{oswald1996a}. Compassion is often described as going beyond this, and suggests concern for the sufferings of others, but also action based on that concern \cite{zaki2017a}\cite{nussbaum1996a}. 

Empathy and compassion may lead to a better understanding of the sufferings of others, and thus efforts to reduce it. One way this works is through perspective-taking \cite{schrier2019designing}\cite{batson1997a}. However, empathy has also been critiqued in terms of its effectiveness in leading to prosocial outcomes \cite{bloom2017a}, as well as the many different ways that people may discuss and measure it \cite{hall2018a}. In this study, we focus on how empathy and compassion may relate to reducing biases related to another ethnic group and perspective-taking, or the ability to take on or consider another’s point of view, feeling, opinion, or experience. 

Games have been used to enhance empathy and compassion, with mixed results \cite{schrier2021a}\cite{schrier2021b}\cite{schrier2019b}. For instance, games have been used to enhance empathy toward other players and/or virtual characters \cite{belman2010a}\cite{schrier2017a}\cite{greitemeyer2010a}\cite{flanagan2014a}\cite{mahood2017a}\cite{harth2017a}. Games have also been critiqued for aiming to enhance empathy, and enhancing empathy is not always beneficial or effective \cite{schrier2021a}\cite{ruberg2020a}. Moreover, games that enhance perspective-taking may not necessarily lead to prosocial outcomes. In one study, Nario-Redmond, Gospodinov, and Cobb used a virtual reality game to share the perspectives of people with a disability. They found that after the game people felt more sad, empathetic, and helpless. While the players were more empathic toward people who are disabled, they did not want to more frequently interact with them, in part because they felt so overwhelmed with sadness once they experienced the experiences of people who are disabled \cite{nario-redmond2017a}.
\subsection{Game jams and empathy}

Less studied have been whether the use of game design (the process of making games) may also enhance empathy and compassion for others. Game jams have been used for learning purposes \cite{aurava2021a}\cite{merilaeinen2020a}, and to support indigenous culture and identity \cite{kultima2019a}, and practice social skill development \cite{smith2016a}. A number of organizations, such as Games for Change and iThrive, have created game jam events for the purpose of supporting social and emotional skills \cite{unknown-a}\cite{ithrive-a}.  Games have also been used explicitly for teaching about social and civic issues \cite{myers2019a}\cite{laumer2020a}. Schrier (2019) used a series of game jams on identity to understand whether the process of making a game would enhance measures of empathy, such as empathic concern, perspective-taking, and feelings of safety. The study found that while participants explored their identity and that of others, and were more open to using games for empathy, they only changed on the feeling of safety measure of empathy. The game jams in this study took place in cities in the United States, such as New York and Seattle, and included mostly game developer professionals and game development students. The game jam study had three different conditions, (One) a use of just a game jam with the theme of “identity,” (Two) a use of a game jam and a guide to support the incorporation of principles into the games, and (Three) a use of the game jam, a guide, and the inclusion of an anti-bias educator as part of the team \cite{schrier2019designing}.

This current study explores the efficacy of a virtual reality game jam in teaching teenagers empathy and compassion. The virtual reality game jam was designed to include conversations on identity, bias, and discrimination with particular focus on ethnicity within the Nigerian context. This current study will fill a gap in our understanding of how game design, and specifically, events like a virtual reality game jam, support the practice of empathy and compassion toward ethnic groups that are not our own. In addition, it will help us to expand our understanding of how game jams can be designed and adopted in African settings like Nigeria.
\section{methodology}
The game jam events were piloted in April of 2021 at Aje Comprehensive Junior and Senior School at Sabo Commercial Avenue, Lagos, with 61 students who are ages 12 to 20. The goal of the game jam was to enhance perspective-taking, identity exploration, and connection to others through the process of design, as well as to encourage students to express themselves through game design. The pilot game jam took place over three different days, and had a theme of “identity.” The first day included set up and establishing the conditions. The second day consisted of a series of identity exercises and brainstorming activities, as well as some instruction on empathy and compassion, as well as some perspective-taking exercises. For instance, one of the activities was “The Story of My Name.” Each group participant needed to spend a few minutes writing a story about their name, and how they received it, and what it means to them (Figure 1). They then took turns reading the story aloud to each other, as a way of sharing their identities and getting to know each other, as well as to help in coming up with ideas for a virtual reality game. By the end of the second day, the participants also started to devise the concept for virtual reality game (Figure 2). The third day of the game jam event consisted of developing the virtual reality game with a team of developers. In total, the pilot lasted around 13 hours and 30 minutes.

There were three conditions in the study. The participants were randomly selected to be in one of the conditions by choosing from among different color papers. In the first condition, the participants did not participate in the game jam at all. In the second condition, the participants participated in days one and two of the game jam (the set up and the brainstorming and conceptualization of the virtual reality game) but not the development of the game. In the third condition, the participants participated in the whole game jam, including the brainstorming and conceptualization, as well as the development of the game. This team included a mentor who helped the students with the game design and development, as none of the students had any game development background. All students participating in the study received a pre-intervention survey and a post-intervention survey. The survey consisted of questions adapted from the fantasy subscale of the Interpersonal Reactivity Scale (IRI) \cite{davis1983a}, a 28-item scale with four subscales; Adolescence Measure of Empathy (AMES) scale, a questionnaire for adolescents that measures affective and cognitive empathy, and sympathy and has 12 total items; Sussex-Oxford Compassion Scale, a 6-item scale that measures whether study participants are giving truthful responses or misrepresenting themselves to present a positive public perception; and Inclusion of Others in Self Scale (IOS), to measure the connectedness of participants with someone in the their group, as well as general demographic questions, and if they were in condition two and three, questions about their experience as part of the game jam and a focus group discussion. The IOS scale was only used in the questionnaire that took place after the intervention. 

\begin{figure}[h]
  \centering
  \graphicspath{ {./images/} }
  \includegraphics[width=\linewidth]{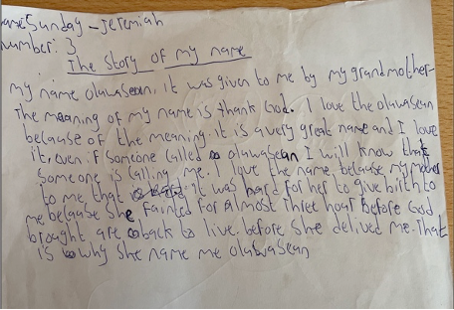}
  \caption{Example of a story written by a participant about their name during the brainstorming and conceptualization exercises.}
  \Description{The story of your name during VR game jam}
\end{figure}

\begin{figure}[h]
  \centering
  \graphicspath{ {./images/} }
  \includegraphics[width=\linewidth]{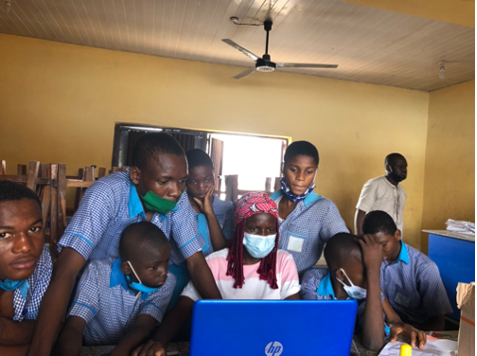}
  \caption{Image of students working together on a game jam game during the pilot study}
  \Description{Students working on the game jam}
\end{figure}
\section{results and next steps}
Across all conditions, 57 participants were in the pilot study. There were 45 males and 12 females, a range of students from 12 to 20. Participants identified as Hausa (n=5; 8.77\%), Yoruba (n=42; 73.68\%), Igbo (n=5; 8.77\%), and Niger Delta (n=5; 8.77\% ). Condition One had 20 participants; Condition Two had 21 and Condition Three had 20 participants recruited for the study. After intervention, the study had an attrition rate of 6.56\% and finally results were computed with 16 participants in Condition one, 21 participants in Condition Two and 15 participants in Condition Three. In Condition One, there were 16 male participants and 4 female participants. In Condition Two, there were 16 male participants and 5 female participants. In Condition Three, there were 13 male participants and 3 female participants.

During the game jam, two games were developed. One was called Forgiveness, which focused on resolving conflicts, understanding that although conflicts can arise, people can learn to forgive others and lend help when they can regardless of the offenses. The second game was called Togetherness, which focused on perspective taking, rejection, tolerance and acceptance. For this game, the player plays as a young boy who was rejected by a community he migrated to after his village got burnt down by a group of people due to ethnic discrimination. So he had to move from village to village seeking for help, the villagers can decide to show kindness or not. Here, the player is meant to take the perspective of the boy, understand his suffering and feel his emotions.

We initially sought to use an ANOVA analysis and to use repeated measures to calculate differences within individuals, and to also compare differences across conditions. This test would inform us if the change in the subjects and between conditions is statistically significant. However, in analyzing the data, we found that the two assumptions of the analysis of variance were not met. The data was not normally distributed for the four scales. We also looked at the homogeneity of the data and for most of the scales the data was also not homogeneous, except for the Sussex scale. We then rejected the data in order to avoid the risk of making a Type 1 error. As a result, our next steps are:

Searching for a nonparametric test to analyze the data.

Conducting another pilot, with more consideration to how we recruit participants and randomize them into different conditions.

Revisiting the scales to see if they are appropriate for the context of a game jam in Nigeria.
\begin{acks}
This work is funded by the Templeton World Charity Foundation. Grant no: TWCF0505. 
\end{acks}
\bibliographystyle{ACM-Reference-Format}
\bibliography{gjparse}
\end{document}